\documentclass{kluwer}    
\let\ni=\noindent
\newcommand{\ea}{{\it et al.}}
\newcommand{\ie}{{\it i.e.\ }}
\newcommand{\fis}{\left (\frac{\dot \phi}{ \phi} \right ) }
\begin{document}                                                                                   
\begin{article}
\begin{opening}         
\title{Isotropic evolution of a JBD anisotropic Bianchi universe} 

\author{ H.\ N.\  \surname{N\'u\~nez-Y\'epez}\email{nyhn@xanum.uam.mx}}
  
\runningauthor{H.\ N.\ N\'u\~nez-Y\'epez}

\runningtitle{Evolution of a JBD   universe}
\institute{Departamento de F\'{\i}sica, Universidad Aut\'onoma Metropolitana-Iztapalapa, Apar\-tado Postal 
21-726, Coyoacan D.\ F.\ CP 04000, M\'exico}
\date{October 22, 2000}

\begin{abstract}
I study the dynamical effects due to the Brans-Dicke scalar $\phi$-field at the early stages of a supposedly anisotropic Universe expansion  in the scalar-tensor cosmology of Jordan-Brans-Dicke. This is done  considering the behaviour of the general solutions for the homogeneous model of Bianchi type VII in the vacuum case. I conclude that the Bianchi-VII$_0$ model shows an isotropic expansion and that its only physical solution is equivalent to a Friedman-Robertson-Walker spacetime whose evolution can, depending on the value of the JBD coupling constant, begin in a singularity and,  after expanding (inflating, if $\omega>0$), shrink to  another, or starting in a non-singular state, collapse to a singularity.  I  also conclude that the general Bianchi-VII$_h$ (with $h\neq 0$) models show strong curvature singularities producing a complete collapse of the homogeinity surfaces to  2D-manifolds, to   1D-manifolds or to  single points. Our analysis depends crucially on the introduction of the so-called intrinsic time, $\Phi$, as the product of the JBD scalar field $\phi$ times a mean scale factor  $a^3=a_1a_2a_3$, in which the finite-cosmological-time evolution of this universe unfolds  into an infinite $\Phi$-range. These universes isotropize from an anisotropic initial state,  thence I conclude that they are stable against anisotropic perturbations.\end{abstract}

\end{opening}           

\section{Introduction}  
                    In the Jordan-Brans-Dicke scalar-tensor theory of gravity (JBD) (Jordan 1959, Brans and Dicke 1961) a massless scalar $\phi$-field is introduced in addition to the pseudo-Riemannian metric of the spacetime $V_4$ occurring in Einstein general relativity (GR); supposedly, this long range field is generated by the whole of matter in the Universe according to Mach's principle (Dicke 1964); furthermore, the inclusion of this field allows Dirac's idea about the secular variation of the gravitational constant $G$ to occur in JBD cosmology through $\phi \sim 1/G$. The scalar field in JBD acts as an additional effective source of $V_4$ geometry, and it is coupled to the tensorial degrees of freedom of the theory by a constant parameter $\omega$ (Ruban and Finkelstein 1975). The value of $\omega$ can be estimated from astronomical observations as $|\omega| \simeq$ 500 to be in accord with current observations. The theory, however,  does not seem to predict anything different from GR with enough supporting observational evidence. 

However and in spite of what  I  have just said, there is a great deal of interest in JBD (and other scalar tensor theories) due to extended inflation and pre-big bang ideas where scalar fields can solve some of the problems of inflation (Steinhard 1993, Gasperini and Veneziano 1994) and may, even, be considered the cause of  it (Cervantes-Cota and Chauvet 1999). Such scalar fiels can be  easily identified with the JBD $\phi$-field; moreover, since the inception of superstring theories which lead naturally to a dilaton theory of gravity with mandatory scalar fields, the importance of JBD theory has been greatly increased.  This comes about since the JBD action functional already includes a string sector where the dilaton field $\phi_D$ can be suitably related with the JBD scalar field as $\phi \propto \exp(-\phi_D)$. The important role of the $\phi$-field of JBD would especially occur at the strongly relativistic stages of the Universe expansion in view of the important role that the so-called scale factor duality plays in the pre-big bang scenario (Gasperini and Veneziano 1994, Clancy \ea\ 1998). Therefore the importance of studying the evolution of  the homogeneous and anisotropic Bianchi universes in JBD cosmology, since the early cosmological expansion of the Universe can be determined by an anisotropic but homogeneous vacuum stage with a non vanishing scalar field. The importance of the scalar field can be ascertained, for example, by the isotropization it is known to cause in the homogeneous but supposedly anisotropic Bianchi universes (N\'u\~nez-Y\'epez 1999, Chauvet and Cervantes-Cota 1995). On the other hand, results obtained in JBD  models can be interpreted as induced gravity cosmological results corresponding to very early epochs of the universe by identifying the JBD scalar field with $2\pi \mbox{\boldmath $\Phi$}^\dagger \mbox{\boldmath $\Phi$}/\omega$ where \mbox{\boldmath $\Phi$} is the SU(5) isotensorial Higgs field and $\omega$ is the JBD coupling constant  and recurring to a spontaneous symmetry breaking mechanism (Cervantes-Cota and Chauvet 1999).          

In this work  I  analyse the early epochs of a supposedly anisotropic Bianchi type VII vacuum universes and show that these exhibit curvature singularities which make their surfaces of homogeinity collapse to 
2D-manifolds, to string-like  1D-manifolds or, even, to single points. Furthermore,   I  show that despite the suppposed anisotropic behaviour of the Bianchi type universes, a Bianchi VII$_0$ universe evolves isotropically and behaves from the start as a Friedman-Roberson-Walker (FRW) universe in spite of possible anisotropic perturbations.  The initial an final states of this VII$_0$ universe depend on the value of the coupling parameter, $\omega$, of the JBD theory; the evolution starts from a singularity, expands  and then reaches another singularity, for $\omega>0$, the expanding phase is inflationary. The coordinates used to perform the analysis, which includes the scaled (by the local volume on the surfaces of homogeneity) $\phi$-field as an intrinsic time coordinate, $\Phi$, are  useful for studying the model evolution until the singularity is reached (N\'u\~nez-Y\'epez 1999). Furthermore, contrarywise to the predictions of General Relativity (GR), this class of cosmological models isotropize as a result of the interaction with the JBD-scalar field not mattering the initial values of the scale factors, therefore,   I  can say that they are  stable under general anisotropic perturbations. 
\section{Bianchi-type VII field equations}

Let us write for the line element of the spacetime, using signatu\-re $+2$  and natural units $c=G=1$, in the so-called sychronous coordinates as

\begin{equation}\label{lineelement}
 ds^2 \, = \, -dt^2 \, + \, h_{ij}(t) \, \mbox{\boldmath $\omega$}^i \, \mbox{\boldmath
$\omega$}^j,
\end{equation}

\noindent
where the $h_{ij}(t)$ is the metric on the surface of homogeinity assumed to depend only on $t$, the synchronous or cosmological time, \mbox{\boldmath $\omega$}$^i$ are the one-forms (Ryan and Shepley 1975) expressing the properties of the 3-surfaces of homogeneity whose specific values, appropriate for the homogeneous but anisotropic Bianchi VII model, are

\begin{subequation}
\begin{equation}
\mbox{\boldmath $\omega$}^1 = a_1 \left( [\eta  - k \nu] dy - \nu dz \right), \end{equation}
\begin{equation}
\mbox{\boldmath $\omega$}^2 = a_2 \left( \nu dy - [\eta  + k \nu] dz \right), \end{equation}
\begin{equation}
\mbox{\boldmath $\omega$}^3 = a_3 dx, 
\end{equation}
\begin{equation}
\mbox{\boldmath $\omega$}^4 = dt,
\end{equation}
\end{subequation}
 
\noindent 
where $\eta \,=\, \exp (-kx) \cos(M x)$, $\nu \,=\, (-M^{-1}) \exp (-kx) \sin(M x)$, $k \,=\, h/2$ and $M \,=\, (1-k^2)^{1/2}\,$. Inserting the line element (1), with the forms (2), into the JBD vacuum field equations (Ruban and Finkelstein 1975),  I  get

\begin{eqnarray}
\lefteqn{\frac{d^{2}} {dt^{2}}(\ln a_i) +
\frac{d} {dt} (\ln a_i) \frac{d} {dt} (\ln a_1 a_2 a_3) +
\frac{d} {dt} (\ln a_i) \fis +{} } 
\nonumber\\ 
& & {} + {\cal A}_i a_1^{-2} + {\cal E}_i \beta_2 + {\cal F}_i \beta_3 = 0, 
\quad i={ 1,2,3}.
\end{eqnarray}

\noindent 
  I  additionally have  the so-called constriction equation, 

\begin{eqnarray} 
\lefteqn{
 \frac{d} {dt}(\ln a_1) \frac{d} {dt}(\ln a_2) +
 \frac{d} {dt}(\ln a_1) \frac{d} {dt}(\ln a_3) +
 \frac{d} {dt}(\ln a_2) \frac{d} {dt}(\ln a_3)  {} }
\nonumber \\ 
& & {}  + \frac{d} {dt}(\ln a_1 a_2 a_3) \fis - \frac{\omega} {2} \fis^2 +
    {\cal A}_4 a_1^{-2} + {\cal E}_4 \beta_2 + {\cal F}_4 \beta_3 = 0,
\end{eqnarray}

\noindent 
this is an equation of the Raychaudhuri type (Raychaudhuri and Modak 1988, Chauvet \ea\ 1992); finally, the scalar field comply with

\begin{equation}
 \frac{d} {dt} \left([a_1 a_2 a_3] \frac{d\phi} {dt} \right) = 0,
\end{equation}

\noindent 
where $\beta_i \equiv (a_i / (2 a_j a_k))^2$ and the indexes $i,j,k$ are to be taken in cyclic order of 1,2,3. Equations (3), (4) and (5) are written in the standard form we introduced  for solving the Bianchi models in JBD (Chauvet \ea\ 1992); the  values for the constants appearing in them are: ${\cal A}_1 = 4 M^2 - (5/2)$, ${\cal A}_2 = {\cal A}_1 - 2 $, ${\cal A}_3 = 0$, ${\cal A}_4 = {\cal A}_1 - 1$, ${\cal E}_1 = {\cal E}_3 = {\cal F}_1 = {\cal F}_2 = -2$, ${\cal E}_2 = {\cal F}_3 = 2$, ${\cal E}_4 = {\cal F}_4 = -1$, and $M^2 = 1 - (h^2 /4)$. Notice that according to these relationships, $h$ must be restricted to be $|h| \leq 2$ (this can also be seen from the differential forms in (2)). Also, the equations for the Bianchi VII$_0$ model can be particularized from the general ones, just by taking $h=0$. Given certain features of the metric and the concourse of the field equations, the model leds  to the following two additional relationships

\begin{equation}
\frac{d} {dt}(\ln a_1) - \frac{d} {dt}(\ln a_2) = 0,   
\end{equation} 

\noindent and

\begin{equation}
\frac{h a_2} {2  a_1^2  a_3}  =  0. 
\end{equation} 

\noindent
Equation (6) always provides an additional relationship between the two scale factors, $a_1$ and $a_2$, not mattering what the value of the $h$-parameter; on the other hand, equation (7) becomes just a trivial identity when $h=0$ not restricting in any way the values of the scale factors. The case $h \neq 0$ and some of its consequences are analysed in subsection 3.2. \par

\subsection{Scaling the $\phi$-field}

From equation (5),  I  easily obtain

\begin{equation}
 a_1 \, a_2 \, a_3 \, \frac{d\phi}{ dt} \,=\, \phi_0,
 \end{equation}

\noindent
where $\phi_0$ is an integration constant; thus  I  can introduce the scaled $\phi$-field, $\Phi$, as $\Phi \,\equiv\, \phi / \phi_0$, which coincides with the so-called intrinsic time (N\'u\~nez-Y\'epez 1995). From equation (8),  I  easily get

\begin{equation}
 \partial_t \,=\, (a_1 \, a_2 \, a_3)^{-1} \, \partial_{\Phi}.
 \end{equation}

\noindent
This equations shows that $\Phi$ is a monotonic function of the synchronous time $t$, $\Phi$ can hence be used also as a time reparametrization useful for solving equations (3). In fact, $\Phi$ has been found useful for analysing the Bianchi vacuum models in several situations (Chauvet \ea\ 1991, 1992, Carretero-Gonz\'alez \ea\ 1994, N\'u\~nez-Y\'epez 1995). For the sake of convenience, let us introduce the notation $(\,)^{\prime} \,\equiv\, \partial_{\Phi}$ and, defining the Hubble expansion rates as $H_i \,\equiv\, (\ln a_i)^{\prime}$, the reparametrized field equations become

\begin{equation}
H_i^{\prime} +\, \frac{H_i}{ \Phi} +\, {\cal J}_i\,a_2^4 \,+\, {\cal K}_i\,a_3^4 \,+\, {\cal N}_i\,a_2^2\, a_3^2 \,=\, 0,
\quad i={ 1,2,3}
  \end{equation}

\noindent
and the constriction equation becomes

\begin{eqnarray}
\lefteqn {
H_1 \, H_2 \,+\, H_1 \, H_3 \,+\, H_2 \, H_3 \,+\, \frac{(\ln a_1 \, a_2 \,
a_3)^{\prime}} {\Phi} \,-\, \frac{\omega} {2 \, \Phi^2} {} } \nonumber\\
& & {} +\, {\cal J}_4 \, a_2^4 \,+\, {\cal K}_4 \, a_3^4 \,+\, {\cal N}_4\,
a_2^2 \, a_3^2 \,=\, 0,
\end{eqnarray}

\noindent
the specific values for the constants appearing in equations (10) and (11) are combinations of the constants previously used: ${\cal J}_1 = {\cal J}_3 = {\cal K}_1 = {\cal K}_2 = -1/2$, ${\cal J}_2 = {\cal K}_3 = 1/2$, ${\cal J}_4 = {\cal K}_4 = -1/4$, ${\cal N}_1 = 4 M^2 - (5/2)$, ${\cal N}_2 = {\cal N}_1 - 2$, ${\cal N}_3 = 0$, ${\cal N}_4 = {\cal N}_1 - 1$. The specific form chosen to write the equations and the parameters just emphasizes the relationship with our previous work (Chauvet \ea\ 1991, 1992, N\'u\~nez-Y\'epez 1995, 1999) on exact solutions for vacuum Bianchi models in JBD.

\section{Solutions for the vacuum Bianchi type VII universes}

For solving the equations of the Bianchi-type VII model,  I  found convenient to address separately the specific VII$_0$ and the generic VII$_h$ Bianchi models. The exact solutions of the next subsections are obtained using essentially the method used in (Chauvet \ea\ 1992).

\subsection{  Bianchi type VII$_0$}

In the reparametrized formulation of the equations for the anisotropic homogeneous metric of Bianchi type VII$_0$, solutions can be obtained for the case of a Bianchi-type VII$_0$; the specific solution depends on the sign of the quantity 

\begin{equation}
\Delta \equiv - 4 ({\cal B} + 1/4), 
\end{equation}

\ni where 

\begin{equation}
{\cal B}=\Phi^2 H_1^2-\frac{\Phi}{ 2} H_1- \frac{\Phi^2}{  2}H_1', 
\end{equation}

\noindent 
is a constant, \ie\ is a first integral of the scaled system (3) and (4), that depends on the Hubble expansion rates. In this way  I  find that out of the possible solutions of equations (10), the only physically plausible is the one corresponding to the case $\Delta < 0$ (the $\Delta > 0$ or $\Delta = 0$ cases can be shown to led to negative or even complex scale factors). For some details see (N\'u\~nez-Y\'epez 1995). The only physically admissible solution can be explicitly written as

\begin{equation}
 a_1(\Phi) = \left( 4\, {\cal B} + 1 \over {c_0}^4 \right)^{1/4} \,
\left( \Phi \cosh \left[ - \sqrt{ ({\cal B} + 1/4)} \ln (f \Phi^2) \right] \right)^{-1/2},
\end{equation}

\noindent
where $c_0$ and $f$ are positive integration constants. The other two scale factors can be easily obtained from $a_1(\Phi)$, as follows from (6) and (10), they are 

\begin{equation}
 a_2(\Phi) = c_0 \, a_1(\Phi) , 
\end{equation}
\begin{equation}
 a_3(\Phi) =\ 2^{-1/2} c_0 a_1(\Phi). 
\end{equation}

\noindent
These results explicitly show that the three scale factors are proportional to each other; this means that the Bianchi-VII$_0$ model, despite what  I  could have anticipated, shows an {\sl isotropic} evolution; it also implies that the shear, vorticity and acceleration of the reference congruence all vanish. The vacuum Bianchi-VII$_0$ JBD universe thus behaves as a Friedmann-Roberson-Walker (FRW) space-time ---in a way, this is not totally surprising since Bianchi cosmologies correspond to the simplest deviations from a FRW environment.  Figure 1 shows the evolution of the scale factors of the model as a function of the intrinsic time (or scaled $\phi$-field) $\Phi$, for the values $\omega=48, 0, -4/3$. The local volume on the surface of homogeneity is then

\begin{equation}
 V \,=\, {c_0^2 \, (a_1)^3 \over \sqrt{2}}.
\end{equation}

In the case $\omega >0$ the model exhibits inflationary behaviour, as can be easily proved from equations (14)--(16), reaches a maximum volume in a finite cosmological time and, in the process, erases any hint of the primordial anisotropies.  It then shrinks again to a singularity (N\'u\~nez-Y\'epez  1999).
The constriction equation implies the following relationship in our case 

\begin{equation}
 12 \, \left( {\cal B} \, + \, {1 \over 4} \right) - (3 \, + \, 2 \, \omega) = 0
\end{equation}

\noindent
which directly implies that ${\cal B}=\omega/6$; from here  I  notice that to have meaningful solutions the coupling parameter has to be restricted to $\omega > -3/2$ or, in terms of ${\cal B}$, the restriction is ${\cal B}> -1/4$. The constriction equation is found thus to depend only on $\omega$,  whose specific value is enough to determine the evolution of the Hubble expansion rates through the nonlinear but elementary equation (13) ---very similar equations are valid in all the integrable Bianchi models (N\'u\~nez-Y\'epez  1995).  I  have also integrated the equation  obtaining (in the case $\omega >0 $)

\begin{equation}
H_1(\Phi)= \sqrt {\cal B} \tanh\left(\hbox{arctanh}(H_0 \Phi_0/\sqrt{\cal B}) + 2\sqrt {\cal B}\log(\Phi_0/\Phi)\right)/\Phi; 
\end{equation}

\noindent
notice that, as the scale factors are proportional to each other, the Hubble rates are all the same at all times: $H_1=H_2=H_3$. From the solution it can be seen that the Hubble expansion rates, after reaching a minimum value, vanish asymptotically as $\sim \log(\Phi)/\Phi$ when $\Phi\to \infty$.  \par

Using equations (9) and (17)  I  can obtain the dependence of $\Phi$ on
$t$, as follows

\begin{equation}\label{t}
 t = { (4 {\cal B} + 1)^{3/4} \over \sqrt{2} \, {c_0}} \int \left(\Phi \cosh [ \sqrt{{\cal B} + {1/ 4}} \, \ln(f \Phi^2)]\right)^{-3/2} d \Phi.
\end{equation}

\noindent
This integral can be explicitly evaluated in terms of the Barnes extended hypergeometric function (Rainville 1980) but, in practical terms, is better to calculate it numerically. It should be noticed that on integrating (\ref{t})  from 0 to $\infty$, the cosmological time reaches a ${\cal B}$-dependent finite value: $t_e$ (N\'u\~nez-Y\'epez 1999);  I  cannot reach beyond this specific value of the cosmological time in our approach. 
Though  I  can obtain $t$ as a function of $\Phi$,   I  cannot invert it to obtain explicitly $\Phi$ as a function of $t$. Nevertheless, figure 2 exhibits the dependence of the field $\Phi$ on $t$, showing the enormous change that occurs in $\Phi$ over a very small span of $t$ values. This also shows that, asymptotically, $\Phi$ grows without bound whereas $t$ approaches a certain finite value $t_{e} (\simeq 3.624$ in this specific case); on the other hand as $t\to 0$, $\Phi$ vanishes. Notice that these conclusions can be applied with no changes to the scalar field $\phi$, excepting when $\phi_0=0$. This behaviour has been instrumental  for explaining, in a Bianchi IX cosmology, the vanishing of the maximum Lyapunov exponent calculated in the intrinsic time $\Phi$ in spite of it being positive in the synchronous time $t$ (Carretero-Gonz\'alez \ea\ 1994). \par

  I  can now relate the behaviour of $\Phi$ with that of $a_1$; since, as  in figure 1 shows, the universe can begin with a singularity and then, as $\Phi$ begins grows, the universe expands  as $\sim \Phi^{\alpha}$, where $\alpha \equiv 3\sqrt{{\cal B} +1/4}-3/2$. But this expansion from a singularity only occurs if $ {\cal B} > 0$ (hence, only if $\omega < 0$), otherwise the universe can start from a non singular state and proceed to collapse to a singularity. In the expansion case, the universe rapidly reaches a maximum volume $V_{\hbox{max}}$, that can be easily calculated from (14) and (17), and then it shrinks  until it reaches again a singularity; see figures 1 and 3 and the comments in section 4. As a conclusion of the previous discussion  I  can say that our chosen time coordinate, $\Phi$, is seem to be  appropriate for studying the early stages of the evolution and certain features of the singularity (see section 4 and figure 2) however, it does not allow us to analyse the behaviour beyond it. However, as I am mainly interested in the early epochs of the universe, this is of no importance.  \par

 \subsection{Bianchi type VII$_h$}

From equations (6) and (7), a few relations can be obtained for the
scale factors irrespective of the value of $h \neq 0$; from (6)  I  get (15) (the same equation than in the case $h=0$). From equation (7),  I  might get $a_2 = 0$, implying that $a_1 = 0$ and that $a_3$ can take any value; then the model collapses into essentially a spatial 
one-dimensional manifold: \ie\ a string-like object. Other possible options allowed by equation (7), imply that $a_2 \, a_3 \to \infty$, thus the spatial 
3-surfaces of homogeinity are seen to collapse into 2-surfaces. The important point here is that for any choice of values for the scale factors in equation (7), the model is found always to spatially collapse. Every Bianchi-type VII$_h$ (with $h \neq 0$) universe is thus highly singular (see section 4). \par

As  I  have been able to get the scale factors for the especific cases addressed in this paper,  I  have obtained exact vacuum solutions for the Bianchi VII models. The important conclusion is that all solutions found  for the generic Bianchi VII model with $h=0$, and $h \neq 0$---with the restriction $|h| \leq 2$--- describe collapsing  universes. The behaviour of the scale factors in the generic cases (with $h= 0$) is exhibitted in figure 1.

\section{Curvature singularities in the Bianchi-type VII universes} 

In this section  I  study the curvature singularities present on the Bianchi-VII universes, though some very specific singularities were discussed in subsection 3.2. \par

 I  say that a universe is singular if the value of the Ricci scalar $R = g^{ab} R_{ab}$ along a congruence of geodesics diverges, $|R| \to \infty$, whereas the associated affine parameter tends to a finite value (Wald 1984). If this happens, then  I  say that there exists a curvature singularity and thus that I am dealing with a singular universe model. Notice also that, due to the choice  I  made for the $\omega$ parameter, here the singularity occurs when $R \to - \infty$ (figure 3) rather than the other way round. \par

 The problem is basically how to choose an appropriate geodesic congruence. For our Bianchi VII models,  I  have chosen as the proper congruences the world lines of test observers (time-like geodesics) whose affine parameter is the synchronous time $t$. Thence, the scalar curvature can readily  be shown to be

\begin{equation}
 R \,=\, 2 \left[ {\ddot a_1 \over a_1} + {\ddot a_2 \over a_2} +
{\ddot a_3 \over a_3} + {\ddot \phi \over \phi} + {\omega \over 2}
\left( {\dot \phi \over \phi}\right)^2 \right].
\end{equation}

\noindent
Using only equations (5), (10) and (11)  I  can rewrite $R$ for the vacuum Bianchi-VII model in terms of the coupling parameter $\omega$ and the scalar field $\phi$, as follows

\begin{equation}
 R \,=\, - \, \omega \, \left( {\dot \phi \over \phi} \right)^2
     \,=\, - \, \omega \, \left( {1 \over a_1 \, a_2 \, a_3 \, \Phi}
\right)^2;
\end{equation}

\noindent
it is important to notice that expressions (20) and (21) do not depend on the $h$-value in any way, they are valid for all the Bianchi-type  models and not only for the type VII I am discussing in this paper (N\'u\~nez-Y\'epez 1995).  I  have expressed the scalar curvature in two different ways, equations (20) or (21), both are important because they exhibit  the explicit dependence of $R$ on the scale factors and $\phi$ and its time derivatives, or on the coupling paramenter $\omega$ and the scale factors (21) and, besides, they have both a certain simplicity. For $R \to \infty$ in (21), all that is needed is that at least one of the scale factors ($a_1$, $a_2$, or $a_3$) or the reescaled scalar field $\Phi$, or just the scalar field $\phi$, vanish at a finite value of the synchronous time $t$. Notice also that  I  can regard the evolution of the Bianchi-type VII universe ---equations (3) and (4)--- as driven by  the curvature $R$, this is especially true near the singularity. \par
 
Notice that the plot of $t$ against $\Phi$ in figure 2, which does not depend on $h$ in any way, shows the rather small range of $\Phi$-values in which the expansion of the model universe VII$_0$ occurs, as  I  can see comparing with figure 1; this also corresponds to the region free of singularities in the model (VII$_0$), as can be seen on comparing with figure 3. On comparing figures 1 and 2, the role of the curvature scalar in governing the expansion can be qualitatively described as follows: a strong curvature prevents expansion; it is only when curvature is small that the full extent of expansion is reached but, as soon as the curvature is large in magnitude again, contraction sets in. This qualitative behaviour is in accord with current ideas (Turner and Weinberg 1997, Clancy \ea\ 1998).

In  the general VII$_h$ JBD model, the universe is always singular, as can be easily seen from equation (7) and the results of section 3.2; in this sense,  I  say that the model VII$_h$ is completely singular.

\section{Concluding remarks} 

The supposed homogeneous and anisotropic Bianchi VII model  shows in fact an isotropic expansion in the case in which $h=0$. On the other hand, this article shows  that the dynamics of the early stages of the expansion in the same JBD model depends on just one of the scale factors (that, here,  I  choose as $a_1$) and that much of the universe evolution  depends on the sign of $\omega$ (or on that of $\cal B$), since equation (13) determines the Hubble expansion rates. This is as happens in other Bianchi models (Chauvet \ea\ 1992).  I  have also obtained the dependence of the three scale factors $a_i$ on the reescaled field $\Phi$. I have concluded that the three scale factors are proportional to each other thence the expansion is not anisotropic, as it is  usually  assumed to be in this model. In fact,  I  have shown that a Bianchi-VII$_0$ JBD vacuum universe is basically equivalent to a FRW-spacetime with no shear, no rotation, and no acceleration; these are characteristics of our actual Universe which can be inferred from the smoothness (and other observational evidence) of the cosmic microwave background radiation.   An important point of the analysis is that it  shows that, even starting with supposedly anisotropic models, the inclusion of a scalar field (the JBD-field) can drastically isotropize the behaviour and thus offers the possibility of coordinating an arbitrary anisotropic vacuum initial state with the observed isotropic properties of the actual Universe ---in spite of the issue of the primordial global anisotropy (Christoulakis \ea\ 2000). This property shows also that the universe dynamics is stable against anisotropic perturbations.
Moreover, this feature does not seem to depend on $\omega$ in any way, it could be valid even with a dynamic coupling parameter as needed for extended inflation models.

The analysis is made possible by the choice of the intrinsic time $\Phi$ as a coordinate.  It should be clear, however, that from such coordinate patch  I  cannot obtain information about what happens with the model in the far  future (\ie\ what happens as $t \to \infty$); from this point of view, the analysis performed refers to the early stages of the model evolution. I have shown that any primordial anisotropy can be erased in a finite cosmological time due to the effect of a scalar field. On the other hand  I  have also shown that certain Bianchi models in the JBD theory can be completely singular, as the type VII$-h$ ($h\neq 0$) analysed here. Notice also that the models show inflationary behaviour when $\omega >0$ and that any primordial anisotropy is smoothed and eventually erased. A possible realistic setting for our model  could be that a primordial vacuum anisotropic  state of the universe  that, due to the presence of a scalar field, appears isotropic and essentially FRW  after, and even during, inflation. 

Nevertheless, for the case of the VII$_h,\, h\neq0$, vacuum model  I  can do conclude that, not mattering the choice   made in equation (7), the universe always collapses. According to the discussion in section 3.2, the homogeneity 3-surfaces of the universe  spatially collapse into  2-surfaces or into  one dimensional objects or, even, into a  single point; this model universe always collapses to a permanent singularity. Such behaviour may make this  model of certain interest if one is just interested in singular behaviour. In this respect, the dependence of the universe dynamics on the curvature scalar is worth pinpointing.

\begin{acknowledgements}
This work has been partially supported by PAPIIT-UNAM through grant IN 122498. It is dedicated with thanks to L.\ Bidsi, M.\ Minina, T.\ Tuga, G.\ Tigga, M.\ Sabi and C.\ F.\ Quimo for all their support and encouragement. It is a pleasure to thank also the colaboration of the colleagues of the Laboratorio de Sistemas Din\'amicos (LSD), UAM-Azcapotzalco. The installations of the LSD where used for the computing needs of this paper. 

\end{acknowledgements}

\eject

{\Large Figure Captions.}
\bigskip

Figure 1. 

The behaviour of the scale factors in a Bianchi-type VII$_0$ universe as a function of the intrinsic time $\Phi$ is shown. The three scale factors are proportional to each other.  I  have used the values
 $f=1$, $c_0=2$, ${\cal B}=\omega/6$, for the parameters in equations (14)--(16).

\ni a) The case $\omega>0$, specifically  I  took $\omega= 48$.

\ni b) The case $\omega=0$.

\ni c) The case $\omega<0$, specifically $\omega=-4/3$.

\bigskip

Figure 2.

The relationship between the cosmological time $t$ and the intrinsic time $\Phi$ showing a monotonic behaviour in every case and the finite range of $t$ values associated with an infinite range in $\Phi$ values. Notice the short span of $\Phi$-values where the main changes in the universe take place. 

\ni a) The case $\omega>0$, specifically $\omega= 48 $.

\ni b) The case $\omega=0$.

\ni c) The case $\omega<0$, specifically $\omega=-4/3$.

\bigskip

Figure 3.

The absolute value of the scalar curvature $R$ in the model dealt with here is shown in the range $0<R<\infty$. I am graphing $\arctan (|R|)$ against $\arctan(\Phi)$. In this way, small $t$-values correspond to small $\Phi$-values, but large values of $\Phi$ correspond to $t\simeq t_e$. The graph was compactified in such a way that a 0 in the horizontal axis is really 0, but when  $\pi/2$ appears, it should be interpreted as $\infty$. The three cases a), b) and c), shown are the same as in figures 1 and 2.

\ni 
\end{article}
\end{document}